\documentclass[a4paper]{report}
\usepackage[utf8]{inputenc}
\usepackage[T1]{fontenc}
\usepackage{RJournal}
\usepackage{amsmath,amssymb,array}
\usepackage{booktabs}
\usepackage{natbib}

\begin{document}

\sectionhead{Contributed research article}
\volume{XX}
\volnumber{YY}
\year{20ZZ}
\month{AAAA}

\begin{article}
\title{hdm: High-Dimensional Metrics}
\author{by Victor Chernozhukov, Chris Hansen, Martin Spindler}

\maketitle

\abstract{
In this article the package High-dimensional Metrics (\CRANpkg{hdm}) is introduced. It is a collection of statistical methods for estimation and quantification of uncertainty in high-dimensional approximately sparse models. It focuses on providing confidence intervals and significance testing for (possibly many) low-dimensional subcomponents of the high-dimensional parameter vector. Efficient estimators and uniformly valid confidence intervals for regression coefficients on target variables (e.g., treatment or policy variable) in a high-dimensional approximately sparse regression model, for average treatment effect (ATE) and average treatment effect for the treated (ATET),  as well for extensions of these parameters to the endogenous setting are provided. 
Theory grounded, data-driven methods for selecting the penalization parameter in Lasso regressions under heteroscedastic and non-Gaussian errors are implemented. Moreover, joint/ simultaneous confidence intervals for regression coefficients of a high-dimensional sparse regression are implemented. Data sets which have been used in the literature and might be useful for classroom demonstration and for testing new estimators are included.}

\section{Introduction}
Analysis of high-dimensional models, models in which the number of parameters to be estimated is large relative to the sample size, is becoming increasingly important. Such models arise naturally in readily available high-dimensional data which have many measured characteristics available per individual observation as in, for example, large survey data sets, scanner data, and text data.  Such models also arise naturally even in data with a small number of measured characteristics in situations where the exact functional form with which the observed variables enter the model is unknown, and we create many technical variables, a dictionary, from the raw characteristics. Examples of this scenario include semiparametric models with nonparametric nuisance functions.  More generally, models with many parameters relative to the sample size often arise when attempting to model complex phenomena.

For \code{R}, many packages for estimation high-dimensional models are already available. For example, \CRANpkg{glmnet} \citep{GLMNet} and \CRANpkg{lars} \citep{larspackage} are popular for Lasso estimation. The section "Regularized \& Shrinkage Methods" in the task view on "Machine Learning and Statistical Learning" contains further implementation of Lasso and related methods.

\nocite{R}

The methods which are implemented in this package (\CRANpkg{hdm}) are chiefly distinct from already available methods in other packages in offering the following four major features: 
\begin{itemize}

\item[\textbf{1)}] First, we provide a version of Lasso regression that expressly handles and allows for non-Gaussian and heteroscedastic errors.

\item[\textbf{2)}] Second, we implement a theoretically grounded, data-driven choice of the penalty level $\lambda$ in the Lasso regressions. To underscore this choice,  we call the Lasso implementation in this package \textquotedblleft rigorous\textquotedblright Lasso (=\code{rlasso}). The prefix \textbf{r} in function names should underscore this. In high-dimensional settings cross-validation is very popular; but it lacks a theoretical justification for use in the present context and some theoretical proposals for the choice of $\lambda$ are often not feasible. Moreover, the theoretically grounded, data-driven choice redundantizes cross-validation which is time-consuming particularly in large data sets.  

\item[\textbf{3)}] Third, we provide efficient estimators and  uniformly valid confidence intervals for various low-dimensional causal/structural parameters  appearing in high-dimensional approximately sparse models.\footnote{A formal definition of approximately sparse models can be found in the accompanying vignette.} For example, we provide efficient estimators and uniformly valid confidence intervals for a regression coefficient on a target variable (e.g., a treatment or policy variable) in a high-dimensional sparse regression model. Target variable in this context means the object not interest, e.g. a prespecified regression coefficient. We also provide estimates and confidence intervals for average treatment effect (ATE) and average treatment effect for the treated (ATET),  as well  extensions of these parameters to the endogenous setting.

\item[\textbf{4)}] Fourth, joint/ simultaneous confidence intervals for estimated coefficients in a  high-dimensional approximately sparse models are provided, based on the methods and theory developed in \cite{BCK2014}. They proposed uniformly valid confidence regions for regressions coefficients in a high-dimensional sparse Z-estimation problems, which include median, mean, and many other regression problems as special cases. In this article we apply this method to the coefficients of a Lasso regression and highlight this method with an empirical example.

\end{itemize}

In the following we will give a short overview over the functionality of the package with simple examples.  A detailed introduction on how to use the functions is given in the accompanying vignette.

\section{A very short guide to the literature}
In this section we give a very short introduction to the literature with a focus on the methods which are implemented in the package.

Lasso under heteroscedastic and non-Gaussian errors was analysed in \citet{BCCH12}. Post-Lasso, i.e. a least squares fit with the variables selected by a Lasso regression, is introduced and analysed in \citet{BC-PostLASSO}. Inference for a low-dimensional component of a potentially high-dimensional vector has been conducted in \citet{BelloniChernozhukovHansen2011}. \citet{BelloniChernozhukovHansen2011} also consider inference on average treatment effects in a heterogeneous treatment effect setting after selection amongst high-dimensional controls. Inference in high-dimensional settings is enabled by the so-called orthogonality condition. The systematic development of the orthogonality condition for inference on low-dimensional parameters in high-dimensional settings is given in \citet{CHS2015}. 

Instrumental variables estimation is a central topic in Econometrics and also becoming more popular in fields such as Biostatistics, Epidemiology, and Sociology. A good introduction to instrumental variables and treatment effects are the books \citet{AngristBook} and \citet{ImbensRubin}. The case of selection on high-dimensional instrumental variables is given in \citet{BCCH12}, the case of selection on the instruments and control variables in \citet{CHS:ManyIVNote}. For further discussion of estimation of treatment effects in a high-dimensional setting including cases with endogenous treatment assignment, we refer to \citet{BCFH:Policy}.

\section{Estimation of Lasso under heteroscedastic and non-Gaussian errors and Inference for low-dimensional subcomponents}
An important feature of our package is that it allows Lasso estimation under heteroscedastic and non-Gaussian errors. This distinguishes the package \CRANpkg{hdm} from other already available software implementations of Lasso. As an additional benefit, the theoretical grounded choice of the penalty does not require cross-validation which might lead to a considerable saving of computation time in many applications.
\subsection{Prediction in Linear Models using Approximate Sparsity}

Consider high dimensional approximately sparse linear regression models. These models have a large number of regressors $p$, possibly much larger than the sample size $n$, but only a relatively small number $s =o(n)$ of these regressors are important for capturing accurately the main features of the regression function. The latter assumption makes it possible to estimate these models effectively by searching for approximately the right set of regressors.

The model reads \[ y_i = x_i' \beta_0  + \varepsilon_i, \quad \mathbb{E}[\varepsilon_i x_i]=0, \quad \beta_0 \in \mathbb{R}^p, 
\quad i=1,\ldots,n \]
where $y_i$ are observations of the response variable, $x_i=(x_{i,j}, \ldots, x_{i,p})$'s are observations of $p-$dimensional  regressors, and $\varepsilon_i$'s are centered disturbances, where possibly $p \gg n$.   Assume that the data sequence is
i.i.d. for the sake of exposition, although the framework covered is considerably more general. An important point is that the errors $\varepsilon_i$ may be non-Gaussian or heteroskedastic \citep{BCCH12}.

The model can be exactly sparse, namely
\[
\| \beta_0\|_0 \leq s = o(n),
\]
or approximately sparse, namely that the values of coefficients, sorted in decreasing
order, $(| \beta_0|_{(j)})_{j=1}^p$ obey,
\[
| \beta_0|_{(j)} \leq \mathsf{A} j^{-\mathsf{a}(\beta_0)},  \quad \mathsf{a}(\beta_0)>1/2, \quad j=1,...,p.
\]
An approximately sparse model can be well-approximated by an exactly sparse model
with sparsity index \[s \propto n^{1/(2 \mathsf{a}(\beta_0))}.\]

In order to get  theoretically justified performance guarantees,  we consider the Lasso 
estimator with data-driven penalty loadings: 
\[ \hat \beta = \arg \min_{\beta \in  \mathbb{R}^p} \mathbb{E}_n [(y_i - x_i' \beta)^2] + \frac{\lambda}{n} ||\hat{\Psi} \beta||_1 \]
where $||\beta||_1=\sum_{j=1}^p |\beta_j|$, $\hat{\Psi}=\mathrm{diag}(\hat{\psi}_1,\ldots,\hat{\psi}_p)$ is a diagonal matrix consisting of  penalty loadings, and $\mathbb{E}_n$ abbreviates the empirical average. The penalty loadings are chosen to insure basic equivariance of coefficient estimates to rescaling of $x_{i,j}$ and can also be chosen to address heteroskedasticity in model errors.   We discuss the choice of $\lambda$ and $\hat \Psi$ below.

Regularization by the $\ell_1$-norm naturally helps the Lasso estimator to avoid overfitting, but it also shrinks the fitted coefficients towards zero, causing a potentially significant bias. In order to remove some of this bias,  consider the Post-Lasso estimator that applies ordinary least squares to the model $\hat{T}$ selected by Lasso, formally, 
\[ \hat{T} = \text{support}(\hat{\beta}) = \{ j \in \{ 1, \ldots,p\}: \lvert \hat{\beta} \rvert >0 \}. \]
The Post-Lasso estimate is then defined as
\[ \tilde{\beta} \in \arg\min_{\beta \in \mathbb{R}^p}  \ \mathbb{E}_n \left( y_i - \sum_{j=1}^p x_{i,j} \beta_j \right) ^2: \beta_j=0 \quad \text{ if } \hat \beta_j = 0 , \quad \forall j. \]
In words, the estimator is ordinary least squares applied to the data after removing the regressors that were not selected by Lasso. The Post-Lasso estimator was introduced and analysed in \citet{BC-PostLASSO}.

A crucial matter is the choice of the penalization parameter $\lambda$.   With the right choice
of the penalty level, Lasso and Post-Lasso estimators possess excellent performance guarantees: They both achieve the near-oracle rate for estimating the regression function, namely with probability $1- \gamma - o(1)$,
\[
\sqrt{\mathbb{E}_n [ (x_{i}'(\hat \beta - \beta_0))^2 ] } \lesssim \sqrt{(s/n) \log p}. 
\]

In high-dimensions setting, cross-validation is very popular in practice but lacks theoretical justification and so may not provide such a performance guarantee. In sharp contrast, the choice of the penalization parameter $\lambda$ in the Lasso and Post-Lasso methods in this package is theoretical grounded and feasible. Therefore we call the resulting method the \textquotedblleft rigorous\textquotedblright  Lasso method and hence add a prefix \textbf{r} to the function names.

\underline{In the case of homoscedasticity}, we set the penalty loadings $\hat{\psi}_j = \sqrt{\mathbb{E}_n x_{i,j}^2}$, which insures basic equivariance properties. There are two choices for penalty level $\lambda$:  the $X$-independent choice
and $X$-dependent choice.  In the $X$-independent choice we set the penalty level
to
\[ \lambda = 2c \sqrt{n} \hat{\sigma} \Phi^{-1}(1-\gamma/(2p)), \]
where $\Phi$ denotes the cumulative standard normal distribution, 
 $\hat \sigma$ is a preliminary estimate of $\sigma = \sqrt{\mathbb{E} \varepsilon^2}$,
and $c$ is a theoretical constant, which is set to $c=1.1$ by default for the Post-Lasso method and $c=.5$ for the Lasso method, and $\gamma$ is the probability level, which is set to $\gamma =.1$ by default.   The parameter $\gamma$ can be interpreted as the probability of mistakenly not removing $X$'s when all of them have zero coefficients.   In the X-dependent case the penalty level is calculated as
\[ \lambda = 2c \hat{\sigma} \Lambda(1-\gamma|X), \]
where
\[ \Lambda(1-\gamma|X)=(1-\gamma)-\text{quantile of}\quad n||\mathbb{E}_n[x_i e_i] ||_{\infty}|X,\]
where $X=[x_1, \ldots, x_n]'$ and $e_i$ are iid $N(0,1)$, generated independently from $X$; this quantity  is approximated by simulation. The $X$-independent penalty is more conservative than the $X$-dependent penalty. In particular the $X$-dependent penalty automatically adapts to highly correlated designs, using less aggressive penalization in this case \citet{BCH2011:InferenceGauss}.

\underline{In the case of heteroskedasticity}, the loadings are set to $\hat{\psi}_j=\sqrt{\mathbb{E}_n[x_{ij}^2 \hat \varepsilon_i^2]}$, where $\hat \varepsilon_i$ are preliminary estimates of the errors. The penalty level
can be $X$-independent \citep{BCCH12}:
\[ \lambda = 2c \sqrt{n} \Phi^{-1} (1-\gamma/(2p)), \]
or it can be X-dependent and estimated by a multiplier bootstrap procedure \citep{CCK:AOS13}:
\[ \lambda = c \times c_W(1-\gamma), \]
where $c_W(1-\gamma)$ is the $1-\gamma$-quantile of the random variable $W$, conditional on the data, where
\[ W:= n \max_{1 \leq j \leq p} |2\mathbb{E}_n [x_{ij} \hat{\varepsilon}_i e_i]|,\]
where $e_i$ are iid standard normal variables distributed independently from the data, and $ \hat{\varepsilon}_i$ denotes an estimate of the residuals.

Estimation proceeds by iteration.  The estimates of residuals $\hat \varepsilon_i$ are initialized by running least squares of $y_i$ on five regressors that are most correlated to $y_i$. This implies conservative starting values for $\lambda$ and the penalty loadings, and leads to the initial Lasso and Post-Lasso estimates, which are then further updated by iteration.  The resulting iterative procedure is fully justified in the theoretical literature.

\subsection{\code{R} Implementation}
The core of the package are the functions \code{rlasso} and \code{rlassoEffects}. They allow estimation of a Lasso regression under heteroscedastic and non-Gaussian errors and inference on a set of prespecified, low-dimensional objects. As Lasso regression introduces shrinkage bias, a post-Lasso option, which can be described as  Lasso estimation followed by a final refit with ols including only the selected variables, is available.

The function \code{rlasso} implements Lasso and post-Lasso. The default option is to use post-Lasso, \code{post=TRUE}. The user can also decide if an unpenalized \code{intercept} should be included (\code{TRUE} by default), \code{LassoShooting.fit} contains the computational algorithm that underlies the estimation procedure. This function implements a version of the Shooting Lasso Algorithm \citep{Fu1998}.  The option \code{penalty} of the function \code{rlasso} allows different choices for the penalization parameter and loadings. It allows for homoscedastic or heteroscedastic errors with default \code{homoscedastic = FALSE}. Moreover, the dependence structure of the design matrix might be taken into consideration for calculation of the penalization parameter with \code{X.dependent.lambda = TRUE}. With the option \code{lambda.start} initial values for the algorithm can be set.\footnote{ In some cases the user might want to use a predefined, fixed penalization parameter. This can be done by a special option in the following way: set homoscedastic to "none" and  supply a value (vector) \code{lambda.start}. Then this value is used as penalty parameter with independent design and heteroscedastic errors to weight the regressors.}

The option \code{penalty} allows to set the constants $c$ and $\gamma$ which are necessary for the calculation of the penalty. For a detailed description how the penalty and variable loadings are calculated, we refer to the accompanying vignette. The maximum number of iterations and the tolerance when the algorithms should stop can be set with \code{control}. 

\code{rlasso} returns an object of S3 class \code{rlasso} for which methods like \code{predict}, \code{print}, \code{summary} are provided.
The methods \code{print} and \code{summary} have the option \code{all}. By setting this option to \code{FALSE} only the coefficients estimated to be non-zero are shown.

The function \code{rlassoEffects} does inference for prespecified target variables. Those can be specified either by the variable names, an integer valued vector giving their position in \code{x}, or by a logical vector indicating the variables for which inference should be conducted. It returns an object of S3 class \code{rlassoEffects} for which the methods \code{summary}, \code{print}, \code{confint}, and \code{plot} are provided. \code{rlassoEffects} is a wrapper function for \code{rlassoEffect} which does inference for a single target regressor.

The function \code{rlassoEffects} might either be used in the form \code{rlassoEffects(x, y, index)} where \code{x} is a matrix, \code{y} denotes the outcome variable and \code{index} specifies the variables of \code{x} for which inference is conducted. This can done by an integer vector (postion of the variables), a logical vector or the name of the variables. An alternative usage is as \code{rlassoEffects(formula, data, I)} where \code{I} is a one-sided formula which specifies the variables for which is inference is conducted. For further details we refer to the help page of the function and the following examples where both methods for usage are shown.

For logistic regression the corresponding functions are implemented in an analogous manner, named \code{rlassologit} and \code{rlassologitEffects}.

\subsection{Example}
First, we generate a data set on which we demonstrate the basic functions:
\begin{example}
set.seed(12345)
> n <- 100 #sample size
> p <- 100 # number of variables
> s <- 3 # number of variables with non-zero coefficients
> X <- matrix(rnorm(n*p), ncol=p)
> beta <- c(rep(5,s), rep(0,p-s))
> Y <- X
\end{example}

Next we use Lasso for fitting and prediction, show the results and make predictions for the old and for new X-variables

\begin{example}
> lasso.reg = rlasso(Y~X, post=FALSE, intercept=TRUE)  # use Lasso, not-post-Lasso
> # lasso.reg = rlasso(X,Y, post=FALSE, intercept=TRUE) # alternative use
> summary(lasso.reg, all=FALSE) 

Call:
rlasso.formula(formula = Y ~ X, post = FALSE, intercept = TRUE)

Post-Lasso Estimation:  FALSE 

Total number of variables: 100
Number of selected variables: 11 

Residuals: 
     Min       1Q   Median       3Q      Max 
-2.09008 -0.45801 -0.01237  0.50291  2.25098 

            Estimate
(Intercept)    0.057
1              4.771
2              4.693
3              4.766
13            -0.045
15            -0.047
16            -0.005
19            -0.092
22            -0.027
40            -0.011
61             0.114
100           -0.025

Residual standard error: 0.8039
Multiple R-squared:  0.9913
Adjusted R-squared:  0.9902
Joint significance test:
 the sup score statistic for joint significance test is 64.02 with a p-value of     0

> yhat.lasso <- predict(lasso.reg)   #in-sample prediction
> Xnew <- matrix(rnorm(n*p), ncol=p)  # new X
> Ynew <-  Xnew
> yhat.lasso.new <- predict(lasso.reg, newdata=Xnew)  #out-of-sample prediction
\end{example}

To reduce bias, now use post-Lasso for fitting and prediction

\begin{example}
> post.lasso.reg = rlasso(Y~X, post=TRUE, intercept=TRUE)
> summary(post.lasso.reg, all=FALSE) 

Call:
rlasso.formula(formula = Y ~ X, post = TRUE, intercept = TRUE)

Post-Lasso Estimation:  TRUE 

Total number of variables: 100
Number of selected variables: 3 

Residuals: 
     Min       1Q   Median       3Q      Max 
-2.83981 -0.80339  0.02063  0.75573  2.30421 

            Estimate
(Intercept)    0.000
1              5.150
2              4.905
3              4.912

Residual standard error: 1.059
Multiple R-squared:  0.9855
Adjusted R-squared:  0.9851
Joint significance test:
 the sup score statistic for joint significance test is 66.87 with a p-value of     0

> yhat.postlasso = predict(post.lasso.reg)  #in-sample prediction
> yhat.postlasso.new = predict(post.lasso.reg, newdata=Xnew)  #out-of-sample prediction
\end{example}

We can do inference on a set of variables of interest, e.g. the first,
second, third, and the fiftieth:
\begin{example}
> lasso.effect = rlassoEffects(x=X, y=Y, index=c(1,2,3,50))
> print(lasso.effect)

Call:
rlassoEffects.default(x = X, y = Y, index = c(1, 2, 3, 50))

Coefficients:
    V1      V2      V3     V50  
5.0890  4.7781  4.8292  0.1384  

> summary(lasso.effect)
[1] "Estimates and significance testing of the effect of target variables"
    Estimate. Std. Error t value Pr(>|t|)    
V1     5.0890     0.1112  45.781   <2e-16 ***
V2     4.7781     0.1318  36.264   <2e-16 ***
V3     4.8292     0.1314  36.752   <2e-16 ***
V50    0.1384     0.1122   1.234    0.217    
---
Signif. codes:  0 ‘***’ 0.001 ‘**’ 0.01 ‘*’ 0.05 ‘.’ 0.1 ‘ ’ 1
\end{example}
The confidence intervals for the coefficients are given by

\begin{example}
> confint(lasso.effect)
         2.5 
> confint(lasso.effect)
         2.5 
V1   4.8710981 5.306834
V2   4.5198436 5.036332
V3   4.5716439 5.086726
V50 -0.0814167 0.358284
\end{example}

Moreover, simultaneous / joint confidence intervals can be calculated which will be discussed in more detail later.

\begin{example}
> confint(lasso.effect, joint = TRUE)
         2.5 
V1   4.8464078 5.3315239
V2   4.4315649 5.1246107
V3   4.4864563 5.1719131
V50 -0.1720149 0.4488822
\end{example} 

Finally, we can also plot the estimated effects with their confidence intervals:
\begin{example}
> plot(lasso.effect, main="Confidence Intervals")
\end{example}

A strength of the package is that it can also handle non-Gaussian and heteroscedastic error what we illustrate with a small example:
\begin{example}
> Y = X
> Y = X
> lasso.reg = rlasso(Y~X, post=FALSE, intercept=TRUE)  # use Lasso, not-post-Lasso
> summary(lasso.reg, all=FALSE) 

Call:
rlasso.formula(formula = Y ~ X, post = FALSE, intercept = TRUE)

Post-Lasso Estimation:  FALSE 

Total number of variables: 100
Number of selected variables: 8 

Residuals: 
     Min       1Q   Median       3Q      Max 
-2.51449 -0.76567  0.09798  0.80007  2.11780 

            Estimate
(Intercept)    0.032
1              4.959
2              4.686
3              4.671
8              0.083
21             0.010
36             0.047
58            -0.070
100           -0.133

Residual standard error: 1.051
Multiple R-squared:  0.9858
Adjusted R-squared:  0.9845
Joint significance test:
 the sup score statistic for joint significance test is 66.87 with a p-value of     0
\end{example}

\section{Joint/ simultaneous confidence intervals}
\subsection{Introduction}
\cite{BCK2014} provide uniformly valid confidence intervals for $p_1$ target parameters which are defined via Huber's Z-problems. The Z-framework encompasses, among other things, mean regression, median regression, generalized linear models, as well as many other methods. The dimension $p_1$ of the target parameter might be potentially much larger than the sample size. Here we apply their results to the Lasso regression which can be embedded into this framework. This enables the provision of uniformly valid joint/ simultaneous confidence intervals for the target parameters. This functionality is implemented via the functions \texttt{rlassoEffects} and \code{confint}. To get joint confidence intervals the option \texttt{joint} of the latter function has to be set to \texttt{TRUE}. In the next section we illustrate this with an empirical illustration, analysing the effect of gender on wage, to quantify potential discrimination.

\subsection{Empirical application: the effect of gender on wage}
In Labour Economics an important question is how the wage is related to the gender of the employed. We use US census data from the year 2012 to analyse the effect of gender and interaction effects of other variables with gender on wage jointly. The dependent variable is the logarithm of the wage, the target variable is \texttt{female} (in combination with other variables). All other variables denote some other socio-economic characteristics, e.g. marital status, education, and experience.  For a detailed description of the variables we refer to the help page.

First, we load and prepare the data.
\begin{example}
> library(hdm)
> data(cps2012)
> X <- model.matrix( ~ -1 + female + female:(widowed + divorced + separated + nevermarried  + 
hsd08+hsd911+ hsg+cg+ad+mw+so+we+exp1+exp2+exp3) + 
+ (widowed + divorced + separated + nevermarried  +
 hsd08+hsd911+ hsg+cg+ad+mw+so+we+exp1+exp2+exp3)^2, data=cps2012)
> dim(X)
[1] 29217   136
> X <- X[,which(apply(X, 2, var)!=0)] # exclude all constant variables
> dim(X)
[1] 29217   116
> index.gender <- grep("female", colnames(X))
> y <- cps2012$lnw
\end{example}

The parameter estimates for the target parameters, i.e. all coefficients related to gender (i.e. by interaction with other variables) are calculated and summarized by the following commands

\begin{example}
> effects.female <- rlassoEffects(x=X, y=y, index=index.gender)
> summary(effects.female)
[1] "Estimates and significance testing of the effect of target variables"
                    Estimate. Std. Error t value Pr(>|t|)    
female              -0.154923   0.050162  -3.088 0.002012 ** 
female:widowed       0.136095   0.090663   1.501 0.133325    
female:divorced      0.136939   0.022182   6.174 6.68e-10 ***
female:separated     0.023303   0.053212   0.438 0.661441    
female:nevermarried  0.186853   0.019942   9.370  < 2e-16 ***
female:hsd08         0.027810   0.120914   0.230 0.818092    
female:hsd911       -0.119335   0.051880  -2.300 0.021435 *  
female:hsg          -0.012890   0.019223  -0.671 0.502518    
female:cg            0.010139   0.018327   0.553 0.580114    
female:ad           -0.030464   0.021806  -1.397 0.162405    
female:mw           -0.001063   0.019192  -0.055 0.955811    
female:so           -0.008183   0.019357  -0.423 0.672468    
female:we           -0.004226   0.021168  -0.200 0.841760    
female:exp1          0.004935   0.007804   0.632 0.527139    
female:exp2         -0.159519   0.045300  -3.521 0.000429 ***
female:exp3          0.038451   0.007861   4.891 1.00e-06 ***
---
Signif. codes:  0 ‘***’ 0.001 ‘**’ 0.01 ‘*’ 0.05 ‘.’ 0.1 ‘ ’ 1
\end{example}

Finally, we estimate and plot confident intervals, first "pointwise" and then the joint confidence intervals.

\begin{example}
> joint.CI <- confint(effects.female, level = 0.95, joint = TRUE)
> joint.CI
                           2.5 
female              -0.244219905 -0.06562666
female:widowed      -0.036833704  0.30902467
female:divorced      0.097104065  0.17677471
female:separated    -0.066454223  0.11305975
female:nevermarried  0.149932792  0.22377417
female:hsd08        -0.230085134  0.28570576
female:hsd911       -0.215291089 -0.02337899
female:hsg          -0.046383541  0.02060398
female:cg           -0.023079946  0.04335705
female:ad           -0.072370082  0.01144259
female:mw           -0.035451308  0.03332443
female:so           -0.043143587  0.02677690
female:we           -0.043587197  0.03513494
female:exp1         -0.008802488  0.01867301
female:exp2         -0.239408202 -0.07963045
female:exp3          0.024602478  0.05229868
> plot(effects.female, joint=TRUE, level=0.95)
\end{example}

This analysis allows a closer look how discrimination according to gender is related to other socio-economic variables.

As a side remark, the version 0.2 allows also now a formula interface for many functions including \texttt{rlassoEffects}. Hence, the analysis could also be done more compact as
\begin{example}
> effects.female <- rlassoEffects(lnw ~ female + female:(widowed + divorced + separated + nevermarried  + 
hsd08+hsd911+ hsg+cg+ad+mw+so+we+exp1+exp2+exp3)+ 
(widowed + divorced + separated + nevermarried  + 
hsd08+hsd911+ hsg+cg+ad+mw+so+we+exp1+exp2+exp3)^2, data=cps2012,
I = ~  female + female:(widowed + divorced + separated + nevermarried  +
 hsd08+hsd911+ hsg+cg+ad+mw+so+we+exp1+exp2+exp3))
\end{example}
The one-sided option \texttt{I} gives the target variables for which inference is conducted.
\section{Data sets}
The package contains six data sets which are made available for researchers. The growth data set contains the so-called Barro-Lee data \citep{BarroLee1994}. It contains macroeconomics information for large set of countries over several decades to analyse the drivers of economic growth. Next, the package includes a data set on settler mortality and economic outcomes which was used in the literature to illustrate the effect of institutions on economic growth \citep{acemoglu:colonial}. The data set \code{pension} contains information about the choice of 401(k) pension plans and savings behaviour in the United States \citep{CH401k}. A data set for analysing the effect of judicial decisions regarding eminent domain on economic outcomes (e.g. house prices), is included. More information is given in \citet{chen:sethi} and \citet{BCCH12}.  The BLP data \citep{BLP} set was analysed in Berry, Levinsohn and Pakes (1995). The data stem from annual issues of the Automotive News Market Data Book. The data set inlcudes information on all models marketed during the the period beginning 1971 and ending in 1990 cotaining 2217 model/years from 997 distinct models. Finally, US census data (CPS2012) is included.\\

All data sets are described in their help pages where also further references can be found.

\section{Further functions for inference on structural parameters}

Additionally, the package contains functions for estimation and inference on structural parameters in a high-dimensional setting. An important tool for estimation in the potential presence of endogeneity is instrumental variable (IV) estimation.  The function \code{rlassoIV} implements methods for IV estimation in a high-dimensional setting. This function is a wrapper for several methods. The user can specify if selection shall be done on the instruments  (z-variables), on the control variables (x-variables), or on both. For the low-dimensional case  where no selection is done for the x- or z-variables \code{rlassoIV} performs classical 2SLS estimation by calling the function \code{tsls}.

The goal of many empirical analyses is to understand and estimate the causal effect of a treatment, e.g. the effect of an endogenous binary treatment $D$, on a outcome, $Y$, in the presence of a binary instrumental variable, $Z$, in settings with very many potential control variables. We provide functions to estimate the local average treatment effect (LATE), the local average treatment effect of the treated (LATET) and as special cases the average treatment effect (ATE) and the average treatment effect of the treated (ATET) in this setting in functions \code{rlassoLATE}, \code{rlassoLATE}, \code{rlassoATE}, and \code{rlassoATET} respectively. These functions also allow calculation of treatment effect standard errors with different bootstrap methods (``wild'', ``normal'', ``Bayes'') besides to the classical plug-in standard errors.
Moreover, methods \code{print}, \code{summary}, and \code{confint}, are available.

Both the family of \code{rlassoIV}-functions and the family of the functions for treatment effects , which are introduced in the next section, allow use with both formula-interface and by handing over the prepard model matrices. Hence the general pattern  for use with formula is \code{function(formula, data, ...)} where formula consists of two-parts and is a member of the class \code{Formula}. These formulas are of the pattern \code{y ~ d + x | x + z} where \code{y} is the outcome variable, \code{x} are exogenous variables, \code{d} endogenous varialbes (if several ones are allowed depends on the concrete function), and \code{z} denote the instrumental variables. A more primitive use of the functions is by simply hand over the required group of variables as matrices: \code{function(x=x, d=d, y=y, z=z)}. In some of the following examples both alternatives are displayed.  

\subsection{Example: IV estimation}
In this section, we briefly present how instrumental variables estimation is conducted in a high-dimensional setting. The eminent domain example, for which the data set is contained in the package, serves as an illustration. The underlying goal is to estimate the effect of pro-plaintiff decisions in cases of
eminent domain (government's takings of private property) on economic outcomes.  The analysis of the effects of such decisions is complicated by the possible endogeneity
between judicial decisions and potential economic outcomes. To address the potential endogeneity, we employ an instrumental
variables strategy based on the random assignment of judges to the federal appellate panels that make the decisions. Because judges are randomly assigned to three-judge panels, the judges and their demographics are randomly assigned conditional on the distribution of characteristics of federal circuit court judges in a given circuit-year.

First, we load the data and construct the matrices with the controls (x), instruments (z), outcome (y), and treatment variables (d). Here we consider regional GDP as the outcome variable. 
\begin{example}
> data(EminentDomain)
> z = EminentDomain$logGDP$z
> x = EminentDomain$logGDP$x
> y = EminentDomain$logGDP$y
> d = EminentDomain$logGDP$d
\end{example}

As mentioned above, $y$ is the economic outcome, the logarithm of the GDP, $d$  the number of pro plaintiff appellate takings decisions in federal circuit court $c$ and year $t$, $x$ is a matrix with control variables, and $z$ is the matrix with instruments. Here we consider socio-economic and demographic characteristics of the judges as instruments.

Next, we estimate the model with selection on the instruments. 
\begin{example}
> lasso.IV.Z = rlassoIV(x=x, d=d, y=y, z=z, select.X=FALSE, select.Z=TRUE) 
> # or lasso.IV.Z = rlassoIVselectZ(x=X, d=d, y=y, z=Z) 
> summary(lasso.IV.Z)
[1] "Estimates and significance testing of the effect of target variables in the IV regression model"
    coeff.     se. t-value p-value
d1 0.01543 0.01926   0.801   0.423
\end{example}

Finally, we do selection on both the $x$ and $z$ variables.
\begin{example}
> lasso.IV.XZ = rlassoIV(x=x, d=d, y=y, z=z, select.X=TRUE, select.Z=TRUE) 
> summary(lasso.IV.XZ)
Estimates and Significance Testing of the effect of target variables in the IV regression model 
     coeff.      se. t-value p-value
d1 -0.03488  0.15881   -0.22   0.826

> confint(lasso.IV.XZ)
        2.5 
d1 -0.3461475 0.2763868
\end{example}

\subsection{Example: Treatment effects}
Here we apply the treatment functions to 401(k) plan participation. 
Though it is clear that 401(k) plans are widely used as vehicles for retirement saving, their effect on assets is less clear. The key problem in determining the effect of participation in 401(k) plans on
accumulated  assets  is  saver  heterogeneity  coupled  with
nonrandom selection into participation states. In particular,
it  is  generally  recognized  that  some  people  have  a  higher
preference for saving than others. Thus, it seems likely that
those individuals with the highest unobserved preference for
saving  would  be  most  likely  to  choose  to  participate  in
tax-advantaged  retirement  savings  plans  and  would  also
have  higher  savings  in  other  assets  than  individuals  with
lower unobserved saving propensity. This implies that conventional estimates that do not allow for saver heterogeneity
and  selection  of  the  participation  state  will  be  biased  upward,  tending  to  overstate  the  actual  savings  effects  of
401(k) and IRA participation.

Again, we start first with the data preparation. For a detailed description of the data set we refer to the help page \code{help(pension)}.
\begin{example}
data(pension)
y = pension$tw; d = pension$p401; z = pension$e401
X = pension[,c("i2", "i3", "i4", "i5", "i6", "i7", "a2", "a3", "a4", "a5",
               "fsize", "hs", "smcol", "col", "marr", "twoearn", "db", "pira", "hown")] # simple model
xvar = c("i2", "i3", "i4", "i5", "i6", "i7", "a2", "a3", "a4", "a5",
         "fsize", "hs", "smcol", "col", "marr", "twoearn", "db", "pira", "hown")
xpart = paste(xvar, collapse = "+")
form = as.formula(paste("tw ~ ", paste(c("p401", xvar), collapse ="+"), "|",  
	paste(xvar, collapse = "+")))
formZ = as.formula(paste("tw ~ ", paste(c("p401", xvar), collapse ="+"), "|", 
	paste(c("e401", xvar), collapse = "+")))
\end{example}

Now we can  compute the estimates 
of the target treatment effect parameters:

\begin{example}
> #pension.ate =  rlassoATE(X,d,y)
> pension.ate = rlassoATE(form, data = pension)
> summary(pension.ate)
Estimation and significance testing of the treatment effect 
Type: ATE 
Bootstrap: not applicable 
   coeff.   se. t-value  p-value    
TE  10490  1920   5.464 4.67e-08 ***
---
Signif. codes:  0 ‘***’ 0.001 ‘**’ 0.01 ‘*’ 0.05 ‘.’ 0.1 ‘ ’ 1


> #pension.atet =  rlassoATET(X,d,y)
> pension.atet = rlassoATET(form, data = pension)
> summary(pension.atet)
Estimation and significance testing of the treatment effect 
Type: ATET 
Bootstrap: not applicable 
   coeff.   se. t-value  p-value    
TE  11810  2844   4.152 3.29e-05 ***
---
Signif. codes:  0 ‘***’ 0.001 ‘**’ 0.01 ‘*’ 0.05 ‘.’ 0.1 ‘ ’ 1


> pension.late = rlassoLATE(X,d,y,z)
> #pension.late = rlassoLATE(formZ, data=pension)
> summary(pension.late)
Estimation and significance testing of the treatment effect 
Type: LATE 
Bootstrap: not applicable 
   coeff.   se. t-value  p-value    
TE  12189  2734   4.458 8.27e-06 ***
---
Signif. codes:  0 ‘***’ 0.001 ‘**’ 0.01 ‘*’ 0.05 ‘.’ 0.1 ‘ ’ 1

> pension.latet = rlassoLATET(X,d,y,z)
> #pension.latet = rlassoLATET(formZ, data=pension)
> summary(pension.latet)
Estimation and significance testing of the treatment effect 
Type: LATET 
Bootstrap: not applicable 
   coeff.   se. t-value p-value    
TE  12687  3590   3.534 0.00041 ***
---
Signif. codes:  0 ‘***’ 0.001 ‘**’ 0.01 ‘*’ 0.05 ‘.’ 0.1 ‘ ’ 1

\end{example}

The results are summarized in Table 1. We see that in this example the estimates are quite similar, while the standard errors differ.

\begin{table}[ht]
\centering
\caption{Estimation of treatment effects}
\begin{tabular}{rrr}
  \hline
 & Estimate & Std. Error \\ 
  \hline
ATE & 10490.07 & 1919.99 \\ 
  ATET  & 11810.45 & 2844.33 \\ 
  LATE & 12188.66 & 2734.12 \\ 
  LATET & 12686.87 & 3590.09 \\ 
   \hline
\end{tabular}
\end{table}

\section{Application: Estimation of the effect of price on demand}
\subsection{Introduction}
A core problems in Economics is to estimate the effect of price on the demand of products. This is usually impeded by the fact that price are not exogenously given, but determined by demand and supply in a simultaneous system of equations, introducing endogeneity. A possible way to solve this problem is to employ instrumental variables - a technique which is also getting more popular outside of Economics. The results found here are a replication in \R of the results in \cite{CHS:ManyIVNote} and \cite{CHS2015}.

here we are interested in estimating the coefficients in a simple logit model of demand for automobiles using market share data. Our example is based on the data and most basic strategy from \cite{BLP}. The goal is to estimate the influence / effect of the price on demand (=market share). Specifically,

$$ \log(s_{it}) - \log(s_{0t}) = \alpha_0 p_{it} + x_{it}' \beta_0 + \varepsilon_{it},$$
$$ p_{it} = z_{it}' \delta_0 + x_{it}' \gamma_0 + u_{it},$$

where $s_it$ is the market share of product $i$ in market $t$ with product zero denoting the outside option, $p_{it}$ is the price and is treated as endogenous, $x_{it}$ are observed included product characteristics, and $z_{it}$ are instruments. In the data set the variable $y$ is defined as  $\log(s_{it}) - \log(s_{0t})$.

In our example, we use the same set of product characteristics ($x$-variables) as used in obtaining the basic results in \cite{BLP}.  Specifically, we use five variables in $x_{it}$: a constant, an air conditioning dummy, horsepower divided by weight, miles per dollar, and vehicle size.  We refer to these five variables as the baseline set of controls.

We also adopt the argument from \cite{BLP} to form our potential instruments.  BLP argue that that characteristics of other products will satisfy an exclusion restriction, $\textrm{E}[\varepsilon_{it}|x_{j\tau}] = 0$ for any $\tau$ and $j \ne i$, and thus that any function of characteristics of other products may be used as instrument for price.  This condition leaves a very high-dimensional set of potential instruments as any combination of functions of $\{x_{j\tau}\}_{j \ne i, \tau \ge 1}$ may be used to instrument for $p_{it}$.  To reduce the dimensionality, BLP use intuition and an exchangeability argument to motivate consideration of a small number of these potential instruments formed by taking sums of product characteristics formed by summing over products excluding product $i$.  Specifically, we form baseline instruments by taking
$$
z_{k,it} = \left(\sum_{r \ne i, r \in \mathcal{I}_f} x_{k,rt} , \sum_{r \ne i, r \notin \mathcal{I}_f} x_{k,rt} \right) 
$$
where $x_{k,it}$ is the $k^{th}$ element of vector $x_{it}$ and $\mathcal{I}_f$ denotes the set of products produced by firm $f$. 
This choice yields a vector $z_{it}$ consisting of 10 instruments.  We refer to this set of instruments as the baseline instruments.

While the choice of the baseline instruments and controls is motivated by good intuition and economic theory, it should be noted that theory does not clearly state which product characteristics or instruments should be used in the model.  Theory also fails to indicate the functional form with which any such variables should enter the model. High-dimensional methods outlined  offer one strategy to help address these concerns which complements the economic intuition motivating the baseline controls and instruments.  As an illustration, we consider an expanded set of controls and instruments.  We augment the set of potential controls with all first order interactions of the baseline variables, quadratics and cubics in all continuous baseline variables, and a time trend which yields a total of 24 $x$-variables.  We refer to these as the augmented controls.  We then take sums of these characteristics as potential instruments following the original strategy which yields 48 potential instruments.  

\subsection{Estimation}
The data set is included in the package an can be accessed in the following way
\begin{example}
data(BLP)
BLPData <- BLP$BLP
\end{example}
A detailed description of the data is given in \cite{BLP} and also on the help page.

In the base line model we consider five x-variables and ten instrumental variables as described above. First we process the data, in particular we construct the design matrices for the x- and z-variables.The matrix of instruments $Z$ is shipped with the data set but could also be constructed with the internal function \texttt{constructIV} in the package \texttt{hdm}.

\begin{example}
attach(BLPData)
X <- as.matrix(cbind(1, BLPData[,c("hpwt", "air", "mpd", "space")]))
Z <- BLP$Z
#Z <- hdm:::constructIV(firm.id, cdid, id, X)
\end{example}

With the baseline x- and z- variables, we estimate the price-effect with ols, tsls, and finally with selection on the x- and z-variables.

\begin{example}
ols.reg <- lm(y~ price + hpwt + air + mpd + space, data =BLPData)
tsls.reg <- tsls(x=X, d=price, y=y, z=Z, intercept =FALSE, homoscedastic = FALSE)
lasso.reg <- rlassoIV(x=X[,-1], y=y, z=Z, d=price, select.X=TRUE, select.Z=TRUE, intercept = TRUE)
\end{example}

The results are summarized in the table below:
\begin{table}[ht]
\centering
\caption{Results baseline model}
\begin{tabular}{rrr}
\hline
Method & Price coefficient & Standard error\\
\hline
Baseline OLS & -0.089 & 0.004\\
Baseline TSLS & -0.136 & 0.012\\
Baseline TSLS with Lasso selection & -0.174 & 0.013\\
\hline
\end{tabular}
\end{table}

Next, we estimate the augmented model which in total has 24 controls and 48 instruments
First, we construct the data needed to estimate the augmented model. The set of augmented IVs can be accessed by

\begin{example}
augZ <- BLP$augZ
\end{example}

\begin{example}
tu <- trend/19
mpdu <- mpd/7
spaceu <- space/2
augX = cbind(1, hpwt, air, mpdu, spaceu, tu, hpwt^2, hpwt^3, mpdu^2, mpdu^3, 
        spaceu^2, spaceu^3, tu^2, tu^3, hpwt*air, mpdu*air, spaceu*air, tu*air,
        hpwt*mpdu, hpwt*spaceu, hpwt*tu, mpdu*spaceu, mpdu*tu, spaceu*tu)
colnames(augX) <- c("constant" , "hpwt" , "air" , "mpdu" , "spaceu" , "tu" , "hpwt^2" , "hpwt^3" , "mpdu^2" ,              "mpdu^3" , "spaceu^2" , "spaceu^3" , "tu^2" , "tu^3" , "hpwt*air" , "mpdu*air" , "spaceu*air" , "tu*air",           "hpwt*mpdu" , "hpwt*spaceu" , "hpwt*tu" , "mpdu*spaceu" , "mpdu*tu" , "spaceu*tu")
# augZ <- hdm:::constructIV(firm.id, cdid, id, augX) # construction of augmented set of IVs
\end{example}

Next, we redo the analysis with augmented set of variables:
\begin{example}
ols.reg <- lm(y~  -1 + cbind(price, augX))
tsls.reg <- tsls(x=augX, d=price, y=y, z=augZ, intercept =FALSE, homoscedastic = FALSE)
lasso.reg <- rlassoIV(x=augX[,-1], y=y, z=augZ, d=price, select.X=TRUE, select.Z=TRUE, intercept = TRUE)
\end{example}

The results for the augmented model are presented in the table:
\begin{table}[ht]
\centering
\caption{Results augmented model}
\begin{tabular}{rrr}
\hline
Method & Price coefficient & Standard error\\
\hline
Augmented OLS & -0.099 & 0.004\\
Augmented TSLS & -0.127 & 0.008\\
Augmented TSLS with Lasso selection & -0.286 & 0.02\\
\hline
\end{tabular}
\end{table}

\subsection{Results}
We report estimation results from the baseline and augmented setting in
the tables above. The estimations all give a negative effect of price on demand as expected, but they differ considerably in size of the effects. In the augmented model we get the strongest influence of price on the market share and we see that the choice of instruments is an important issue.

\section{Summary}
The package \code{hdm} contains methods for estimation and inference in a high-dimensional setting. We have presented a short overview of the functions implemented. The package also contains data sets which might be useful for classroom presentations and as applications for newly developed estimators. More examples and a short introduction in the underlying theoretical concepts are provided in the accompanying vignette. It is planned to extend the functionality and to improve the performance of the functions in subsequent versions of the package. 

\bibliography{chernozhukov-hansen-spindler}

\begin{thebibliography}{20}
\providecommand{\natexlab}[1]{#1}
\providecommand{\url}[1]{\texttt{#1}}
\expandafter\ifx\csname urlstyle\endcsname\relax
  \providecommand{\doi}[1]{doi: #1}\else
  \providecommand{\doi}{doi: \begingroup \urlstyle{rm}\Url}\fi

\bibitem[Acemoglu et~al.(2001)Acemoglu, Johnson, and
  Robinson]{acemoglu:colonial}
D.~Acemoglu, S.~Johnson, and J.~A. Robinson.
\newblock The colonial origins of comparative development: An empirical
  investigation.
\newblock \emph{American Economic Review}, 91\penalty0 (5):\penalty0
  1369--1401, 2001.

\bibitem[Angrist and Pischke(2008)]{AngristBook}
J.~D. Angrist and J.-S. Pischke.
\newblock \emph{Mostly Harmless Econometrics: An Empiricist's Companion}.
\newblock Princeton University Press, 2008.

\bibitem[Barro and Lee(1994)]{BarroLee1994}
R.~J. Barro and J.-W. Lee.
\newblock Data set for a panel of 139 countries.
\newblock \emph{NBER, http://www.nber.org/pub/barro.lee.html}, 1994.

\bibitem[Belloni and Chernozhukov(2013)]{BC-PostLASSO}
A.~Belloni and V.~Chernozhukov.
\newblock Least squares after model selection in high-dimensional sparse
  models.
\newblock \emph{Bernoulli}, 19\penalty0 (2):\penalty0 521--547, 2013.
\newblock ArXiv, 2009.

\bibitem[Belloni et~al.(2010)Belloni, Chernozhukov, and
  Hansen]{BCH2011:InferenceGauss}
A.~Belloni, V.~Chernozhukov, and C.~Hansen.
\newblock Inference for high-dimensional sparse econometric models.
\newblock \emph{Advances in Economics and Econometrics. 10th World Congress of
  Econometric Society. August 2010}, III:\penalty0 245--295, 2010.
\newblock ArXiv, 2011.

\bibitem[Belloni et~al.(2012)Belloni, Chen, Chernozhukov, and Hansen]{BCCH12}
A.~Belloni, D.~Chen, V.~Chernozhukov, and C.~Hansen.
\newblock Sparse models and methods for optimal instruments with an application
  to eminent domain.
\newblock \emph{Econometrica}, 80:\penalty0 2369--2429, 2012.
\newblock Arxiv, 2010.

\bibitem[Belloni et~al.(2013)Belloni, Chernozhukov, Fern\'andez-Val, and
  Hansen]{BCFH:Policy}
A.~Belloni, V.~Chernozhukov, I.~Fern\'andez-Val, and C.~Hansen.
\newblock Program evaluation with high-dimensional data.
\newblock \emph{arXiv:1311.2645}, 2013.
\newblock ArXiv, 2013.

\bibitem[Belloni et~al.(2014{\natexlab{a}})Belloni, Chernozhukov, and
  Hansen]{BelloniChernozhukovHansen2011}
A.~Belloni, V.~Chernozhukov, and C.~Hansen.
\newblock Inference on treatment effects after selection amongst
  high-dimensional controls.
\newblock \emph{Review of Economic Studies}, 81:\penalty0 608--650,
  2014{\natexlab{a}}.
\newblock ArXiv, 2011.

\bibitem[Belloni et~al.(2014{\natexlab{b}})Belloni, Chernozhukov, and
  Kato]{BCK2014}
A.~Belloni, V.~Chernozhukov, and K.~Kato.
\newblock Uniform post-selection inference for least absolute deviation
  regression and other z-estimation problems.
\newblock \emph{Biometrika}, 2014{\natexlab{b}}.
\newblock \doi{10.1093/biomet/asu056}.

\bibitem[Berry et~al.(1995)Berry, Levinsohn, and Pakes]{BLP}
S.~Berry, J.~Levinsohn, and A.~Pakes.
\newblock Automobile prices in market equilibrium.
\newblock \emph{Econometrica}, 63:\penalty0 841--890, 1995.

\bibitem[Chen and Sethi(2010)]{chen:sethi}
D.~L. Chen and J.~Sethi.
\newblock Does forbidding sexual harassment exacerbate gender inequality.
\newblock unpublished manuscript, 2010.

\bibitem[Chernozhukov and Hansen(2004)]{CH401k}
V.~Chernozhukov and C.~Hansen.
\newblock The impact of 401(k) participation on the wealth distribution: An
  instrumental quantile regression analysis.
\newblock \emph{Review of Economics and Statistics}, 86\penalty0 (3):\penalty0
  735--751, 2004.

\bibitem[{Chernozhukov} et~al.(2013){Chernozhukov}, {Chetverikov}, and
  {Kato}]{CCK:AOS13}
V.~{Chernozhukov}, D.~{Chetverikov}, and K.~{Kato}.
\newblock Gaussian approximations and multiplier bootstrap for maxima of sums
  of high-dimensional random vectors.
\newblock \emph{Annals of Statistics}, 41:\penalty0 2786--2819, 2013.

\bibitem[Chernozhukov et~al.(2015{\natexlab{a}})Chernozhukov, Hansen, and
  Spindler]{CHS2015}
V.~Chernozhukov, C.~Hansen, and M.~Spindler.
\newblock Valid post-selection and post-regularization inference: An
  elementary, general approach.
\newblock \emph{Annual Review of Economics}, 7\penalty0 (1):\penalty0 649--688,
  2015{\natexlab{a}}.
\newblock \doi{10.1146/annurev-economics-012315-015826}.

\bibitem[Chernozhukov et~al.(2015{\natexlab{b}})Chernozhukov, Hansen, and
  Spindler]{CHS:ManyIVNote}
V.~Chernozhukov, C.~Hansen, and M.~Spindler.
\newblock Valid post-selection and post-regularization inference in linear
  models with many controls and instruments.
\newblock \emph{American Economic Review: Papers and Proceedings},
  2015{\natexlab{b}}.

\bibitem[Friedman et~al.(2010)Friedman, Hastie, and Tibshirani]{GLMNet}
J.~Friedman, T.~Hastie, and R.~Tibshirani.
\newblock Regularization paths for generalized linear models via coordinate
  descent.
\newblock \emph{Journal of Statistical Software}, 33\penalty0 (1):\penalty0
  1--22, 2010.

\bibitem[Fu(1998)]{Fu1998}
W.~J. Fu.
\newblock Penalized regressions: The bridge versus the lasso.
\newblock \emph{Journal of Computational and Graphical Statistics}, 7\penalty0
  (3):\penalty0 397--416, 1998.
\newblock \doi{10.1080/10618600.1998.10474784}.

\bibitem[Hastie and Efron(2013)]{larspackage}
T.~Hastie and B.~Efron.
\newblock \emph{lars: Least Angle Regression, Lasso and Forward Stagewise},
  2013.
\newblock URL \url{https://CRAN.R-project.org/package=lars}.
\newblock R package version 1.2.

\bibitem[Imbens and Rubin(2015)]{ImbensRubin}
G.~W. Imbens and D.~B. Rubin.
\newblock \emph{Causal Inference for Statistics, Social, and Biomedical
  Sciences: An Introduction}.
\newblock Cambridge University Press, New York, NY, USA, 2015.
\newblock ISBN 0521885884, 9780521885881.

\bibitem[{R Core Team}(2012)]{R}
{R Core Team}.
\newblock \emph{R: A Language and Environment for Statistical Computing}.
\newblock R Foundation for Statistical Computing, Vienna, Austria, 2012.
\newblock URL \url{http://www.R-project.org/}.
\newblock {ISBN} 3-900051-07-0.

\end{thebibliography}
\newpage
\address{Victor Chernozhukov\\
  Massachusetts Institute of Technology\\
	Economics Department and Center for Statistics\\
  50 Memorial Drive, Cambridge, MA 02142\\
  USA\\}
\email{vchern@mit.edu}

\address{Chris Hansen\\
  University of Chicago\\
Booth School of Business\\
5807 South Woodlawn Avenue, Chicago, Illinois 60637\\
  USA\\}
\email{chansen1@chicagobooth.edu}

\address{Martin Spindler\\
  University of Hamburg and Max Planck Society\\
  Von-Melle-Park 5\\
20146 Hamburg\\
  Germany\\}
\email{spindler@mea.mpisoc.mpg.de}

\end{article}

\end{document}